\documentclass{iopjournal}

\usepackage{lineno,tikz, xcolor, amsmath}
\usepackage[square,sort&compress,numbers]{natbib}
\usepackage{braket, graphicx}
\newcommand{\Tr}{\mbox{\rm Tr}}
\usetikzlibrary{arrows,shapes, positioning,shadows,svg.path,backgrounds,fit}

\begin{document}

\articletype{paper}

\title{Feasibility of entanglement-based QKD protocols with SPDC and QD sources}

\author{Mariia Gumberidze$^{1,*}$\orcid{0000-0002-0101-7221}, Vladyslav Usenko$^2$\orcid{0000-0002-8765-8758}}

\affil{$^1$Department of Optics, Palacky University, Olomouc, Czechia}

\affil{$^*$Author to whom any correspondence should be addressed.}

\email{gumberidze@optics.upol.cz}


\begin{abstract}
We theoretically analyze the feasibility of entanglement-based quantum key distribution (QKD) protocols considering widely used spontaneous parametric down-conversion (SPDC) and novel quantum dot (QD) sources. We account for multiphoton emission in SPDC sources and fine-structure splitting (FSS) in QD. In addition, we incorporate imperfect detection, including dark counts and limited efficiency. For SPDC sources, we confirm that the presence of vacuum and multiphoton pairs renders them unsuitable for secure device-independent (DI) QKD implementations under standard detection strategies. Conversely, in the case of QD sources, accounting for the effects of FSS, results in reduced performance of protocols. Our findings are crucial for the practical implementation of entanglement-based QKD protocols using realistic sources and detectors.
\end{abstract}

\section{Introduction}

Quantum key distribution \cite{Pirandola:20} is a quantum technology aimed at distributing secret cryptographic keys between two spatially distant trusted parties. The principles of quantum physics then provide provable security of the keys. QKD was first suggested based on polarized single photons in the celebrated BB84 protocol \cite{ieee1984proceedings} with security arguments based on the no-cloning theorem \cite{Wootters1982}, which prevents perfect copying of unknown quantum states by a potential eavesdropper. Later, it was shown that QKD can be realized using entangled photon pairs in the E91 protocol \cite{Ekert1991} based on the Bell inequality violation, which indicates that no eavesdropper tampered with an entangled EPR state (named after the famous Einstein-Podolsky-Rosen paradox \cite{PhysRev.47.777}). It was then shown that BB84 can be equivalently realized using entangled states \cite{PhysRevLett.68.557}, while the relation to Bell inequality violation allowed for the development a stricter, DI approach to the security of QKD. In DI-QKD, contrary to conventional QKD protocols like BB84 and its modifications, no assumptions are made on the internal workings of the devices, and loophole-free violation of local realism allows to verify the security of the distributed keys.  This enables QKD immune to quantum hacking attacks \cite{RevModPhys.92.025002}, which exploit practical device imperfections, deviating from theoretical models of conventional QKD protocols \cite{Scarani2008}.
DI-QKD protocol was proposed by Acín et al. in \cite{Acin2007PRL} based on E91, incorporating the security analysis of high-dimensional states \cite{AcinPRL2006} and asymmetric basis selection and refined error analysis techniques \cite{Lo2005}. Several DI-QKD variants exist (see \cite{Primaatmaja2023securityofdevice, Zapatero2023-qj} for the recent reviews of DI-QKD), with the standard approach relying on the Clauser-Horne-Shimony-Holt (CHSH) Bell inequality \cite{Brunner2014, PhysRevLett.23.880}. Other versions employ generalized CHSH inequalities \cite{Sekatski2021deviceindependent}, asymmetric CHSH \cite{Woodhead2021}, noisy preprocessing \cite{PhysRevLett.124.230502, PhysRevA.110.042403}, or random key bases \cite{Schwonnek2021-iw}, aiming to enhance protocol robustness. The key performance parameters for these protocols are critical detection efficiency and quantum bit error rate (QBER) under a depolarizing noise model. 

DI-QKD has been explored both theoretically and experimentally using single-photon \cite{Gonzalez-Ruiz:24, Kolodynski2020deviceindependent} and entangled photon states \cite{PhysRevLett.129.050502}. Proof-of-concept experiments have also been conducted using alternative platforms, such as trapped ions \cite{Nadlinger2022-ze} and atoms \cite{PhysRevLett.128.110506}. Similarly, the conventional BB84 protocol and its modifications were studied and demonstrated using weak coherent pulses \cite{Bennett1992} or using entangled photon pairs, as we discuss below. Importantly, the entanglement-based approach not only enables realizations of QKD protocols, including DI-QKD, but can be upscaled using, e.g. entanglement swapping \cite{PhysRevLett.71.4287} towards future quantum networks \cite{wehner2018quantum}. Most of the entanglement-based tests and realizations of QKD were performed using SPDC sources, based on the conversion of pump photons into photon pairs in a nonlinear crystal \cite{Couteau03072018}. All DI-QKD implementations rely on loophole-free Bell tests, with SPDC-based experiments typically utilizing the CHSH inequality \cite{Vivoli2015, Tsujimoto2018}. In addition, SPDC sources have been widely used in conventional QKD implementations, both in entanglement-based configurations \cite{Poppe:04} and in the heralded single-photon regime \cite{PhysRevLett.84.4729}. Alternatively, entangled photons can be generated using QD sources, where conversion occurs in a semiconductor nanoparticle \cite{GarciaDeArquer2021}. Recently, QD sources were used to demonstrate DI-QKD in heralded single-photon \cite{Gonzalez-Ruiz:24}, and entanglement-based realizations \cite{Basset2021}. Conventional entanglement-based QKD protocols were similarly demonstrated using QD sources \cite{Basset2021, BassoBasset2023}. While SPDC sources are broadly used for entanglement-based QKD and QD sources are gaining popularity, it is essential to study the feasibility of QKD with both types of sources under realistic conditions.

In this paper, we study the feasibility of DI-QKD and entanglement-based BB84 protocols, shown in Fig.~\ref{plot1}, using realistic entangled photon sources, specifically SPDC and QDs, under practical conditions. We evaluate the Devetak-Winter bound on the key rate \cite{Acin2007PRL, Pironio_2009} under collective eavesdropping attacks and assess the robustness of DI-QKD to practical imperfections via Bell inequality violation \cite{CHSH}. Unlike prior studies that employed state tomography to analyze SPDC sources \cite{Hosak}, we adopt a photodetection-theory-based approach \cite{Semenov2009, Gumberidze2016, bookWogel} to estimate Bell parameter and QBER, allowing for a direct comparison of these two source types within a unified framework. Additionally, we examine the effects of FSS in QD sources, which introduce phase shifts in Bell states. In contrast to previous comparisons that assumed idealized QD sources and low-gain SPDC, we perform a comprehensive feasibility study incorporating realistic imperfections of sources and detectors, including multi-photon emission in SPDC, FSS in QDs, limited detection efficiency, dark counts, and channel depolarization. The paper is organised as follows: In Section~\ref{methods}, we derive the secret-key expressions within a photodetection-theory framework and model SPDC and QD sources, accounting for realistic imperfections. In Section~\ref{resultsDI}, we evaluate and compare the practical performance of both sources. In Section~\ref{concls}, we draw the main conclusions and sketch the avenues for future research.

\section{Protocols and methods}
\label{methods}

In DI-QKD, Alice and Bob do not trust their source or measurement devices, as imperfections or potential manipulation may alter measurement outcomes. Moreover, they cannot assume a fixed Hilbert space dimension and must analyze the worst-case scenario over arbitrary-dimensional spaces to bound Eve’s information using obtained data. In contrast, conventional QKD such as BB$84$ relies on well-characterized measurements and fixed-dimensional Hilbert spaces, restricting Eve’s knowledge based on directly observed data.

We study the feasibility of entanglement-based DI-QKD and BB84 with realistic entangled photon sources and single-photon detectors. To perform fair comparison between different types of sources, we exclude additional protocol modifications and focus on comparing performance with standard protocol settings, incorporating detection imperfections and assuming depolarization in the quantum channel.

\subsection{Secure key rates}
\label{section-1}
\begin{figure}[t]
  \centering \includegraphics[width=0.7\linewidth]{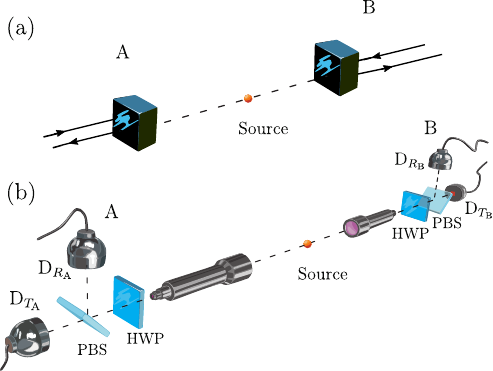}
\caption{(a) Generic scheme of a DI-QKD protocol, adapted from \cite{Acin2007PRL, Pironio_2009}. DI-QKD is designed to operate without assumptions about the internal working of measurement devices, which are treated as black boxes with multiple inputs and outputs. These devices, along with the source of entangled photons, are considered untrusted and potentially controlled by an eavesdropper, making DI-QKD immune to eavesdropping attacks targeting device imperfections. (b) Realistic setup for implementation of entanglement-based BB84 and DI-QKD using polarization analyzers on each side (A and B). The analyzers consist of a half-wave plate (HWP), polarizing beam splitters (PBS), and two detectors (denoted $D_R$ and $D_T$, respectively, for the light reflected or transmitted by a PBS).}
  \label{plot1}
\end{figure}

We study the most common entanglement-based QKD protocols based on the CHSH Bell inequality by evaluating the Devetak-Winter bound on secret key rate, which assumes collective eavesdropping attacks \cite{Acin2007PRL}. The bound for the DI-QKD protocol reads
\begin{equation}  \label{rdw1}
    \displaystyle{r_{DW}^{\textit{DI}} \geq 1-h(Q)-h\left(\frac{1+\sqrt{(S/2)^{2}-1}}{2}\right),}
\end{equation}
where $h(Q)=-(1-Q)\log_{2}(1-Q)-Q\log_{2}(Q)$ is the binary entropy function (hence, the rate is evaluated in bits per channel use), $Q$ stands for QBER and $S$ is the Bell parameter.

For the entanglement-based BB84 protocol, we will use the key rate evaluated in terms of QBER. Assuming a symmetric channel, \(Q_x=Q_z\equiv Q\), the standart proof \cite{Scarani2008} yields
\begin{equation}
  r_{\text{DW}}^{\textit{BB$84$}}
  \;=\;
  1 - 2\,h \bigl(Q\bigr).
  \label{eq:DW_Q}
\end{equation} 

Let us assume we have the maximally entangled Bell state \( \ket{\Phi^{+}} \), defined as:

\[
\ket{\Phi^{+}} = \frac{1}{\sqrt{2}}(\ket{00}+\ket{11}).
\]

The considered QKD scenario involves three possible measurement bases on Alice's side, corresponding to the measurement settings \( A_0, A_1, A_2 \), and two bases on Bob's side corresponding to settings \( B_1, B_2 \) \cite{Pironio_2009}.

We evaluate the CHSH Bell parameter using the following form:
\begin{equation}
    S = \langle a_1 b_1 \rangle + \langle a_1 b_2 \rangle - \langle a_2 b_1 \rangle + \langle a_2 b_2 \rangle,
\end{equation}
where the expectation values for the maximally entangled state \(\ket{\Phi^{+}}\) are given by:
\begin{equation}
    \langle a_i b_j \rangle = \Tr\left[\ket{\Phi^{+}}\!\bra{\Phi^{+}} \hat{A}_i \otimes \hat{B}_j\right], \quad i,j=\{1,2\}.
\end{equation}

The QBER is evaluated using Alice's third measurement setting \(A_0\) and Bob’s first measurement setting \(B_1\). Explicitly, the QBER is given by:
\begin{equation}
    Q = \bra{01}\hat{A}_0 \otimes \hat{B}_1\ket{01} + \bra{10}\hat{A}_0 \otimes \hat{B}_1\ket{10}.
\end{equation}

This formulation clearly separates the role of the third measurement basis on Alice's side, used solely to quantify QBER, from the first two bases that are employed to evaluate the CHSH Bell parameter for security analysis.

The optimal measurement settings for Alice and Bob, expressed in terms of Pauli matrices, are given by:
$$\hat{\mathit{A}}_{0}=\hat{\mathit{B}}_{1}=\hat{\sigma}_{z}, \,\hat{\mathit{B}}_{2}=\hat{\sigma}_{x}, \,\hat{\mathit{A}}_{1}=(\hat{\sigma_{x}}+\hat{\sigma}_{z})/\sqrt{2}, \,\hat{\mathit{A}}_{2}=(\hat{\sigma}_{x}-\hat{\sigma}_{z})/\sqrt{2}.$$
These measurement settings are optimal not only for the Bell state \(\ket{\Phi^{+}}\), but also for the singlet state \(\ket{\Psi^{-}}\), the prototypical entangled state generated via SPDC \cite{Ekert1991}.

Our objective is to verify the validity of this result while accounting for imperfections arising from the specific characteristics of the source and the detailed aspects of the detection process.

\subsection{Photodetection theory}
\label{section-2}

We begin by introducing detector imperfections \cite{bookWogel, Semenov2010}, particularly for single-photon avalanche diode (SPAD) detectors . We describe these imperfections using positive operator-valued measures (POVMs) \cite{Gumberidze2016}. The POVM elements for the transmitted (T) and reflected (R) output ports of a polarization beam splitter (PBS) on Alice's (A) or Bob's (B) side are represented as \( i_{A(B)} \), with \( i \in \{T,R\} \). Each detector has two POVM outcomes: \(0\) indicating no detection event and \(c\) indicating a detection event, given explicitly by:
\[
\hat{\Pi}_{i_{A(B)}}^{(0)} = :\exp\left(-\eta \, \hat{a}^\dagger_{i_{A(B)}} \hat{a}_{i_{A(B)}} - \nu\right):,
\]
\[
\hat{\Pi}_{i_{A(B)}}^{(c)} = 1 - :\exp\left(-\eta \, \hat{a}^\dagger_{i_{A(B)}} \hat{a}_{i_{A(B)}} - \nu\right):,
\]
where \(\hat{a}_{i_{A(B)}}\) and \(\hat{a}^\dagger_{i_{A(B)}}\) are the annihilation and creation operators, respectively, for the optical field modes at the PBS output port \( i_{A(B)} \). The parameter \(\eta\) represents the detection efficiency, and \(\nu\) denotes the detector dark count rate. The notation \(:\ldots:\) indicates normal ordering of operators.

The joint probabilities of photon detection \cite{Semenov2010, Gumberidze2016} are
\begin{equation}
P_{i_{A} i_{B}}\left(\theta_{A}, \theta_{B}\right)
= \mathrm{Tr}\left(
\hat{\Pi}_{i_{A}}^{(c)} \hat{\Pi}_{i_{B}}^{(c)}
\hat{\Pi}_{j_{A}}^{(0)} \hat{\Pi}_{j_{B}}^{(0)}
\hat{\rho}\right),
\label{ProbabilitySquash}
\end{equation}
where $\hat{\rho}$ is the density matrix of the incident light, and 
\(i_{A(B)} \in \{T_{A(B)}, R_{A(B)}\}\) denote the transmitted (\(T\)) or reflected (\(R\)) detection outcomes at Alice’s (A) or Bob’s (B) side, with the condition \(i_{A(B)} \neq j_{A(B)}\).
 The correlation coefficients are then defined as
\begin{equation}
E\left(\theta_{A}, \theta_{B}\right) =
\frac{P_{\text{sam}}\left(\theta_{A},
\theta_{B}\right)-P_{\text{diff}}\left(\theta_{A},
\theta_{B}\right)}{P_{\text{sam}}\left(\theta_{A},
\theta_{B}\right)+P_{\text{diff}}\left(\theta_{A},
\theta_{B}\right)},\label{correlation}
\end{equation}
where \( P_\mathrm{sam} \) and \( P_\mathrm{diff} \) represent the probabilities of same and different events and are defined as:
\begin{align}
P_{\text{sam}}\!\left(\theta_{A}, \theta_{B}\right) &= P_{T_{A} T_{B}}\!\left(\theta_{A},
\theta_{B}\right) + P_{R_{A}
R_{B}}\!\left(\theta_{A},
\theta_{B}\right),\\
P_{\text{diff}}\!\left(\theta_{A}, \theta_{B}\right) &= P_{T_{A} R_{B}}\!\left(\theta_{A},
\theta_{B}\right) + P_{R_{A}
T_{B}}\!\left(\theta_{A}, \theta_{B}\right).
\end{align}

The Bell parameter for the two sets of polarization angles reads 
\begin{equation}
S =\left|E\left(\theta_{A}^{(1)},
\theta_{B}^{(1)}\right)+E\left(\theta_{A}^{(1)},
\theta_{B}^{(2)}\right)-E\left(\theta_{A}^{(2)},\theta_{B}^{(1)}\right)+E\left(\theta_{A}^{(2)},
\theta_{B}^{(2)}\right)\right|.  
\label{bell}
\end{equation}

The next step is to define the angles that correspond to the Pauli matrices in the ideal scenario.

\subsection{Measurements}
\label{section-3}
The polarization analyzer setups on both sides, shown in Fig.~\ref{plot1} (b), include half-wave plates (HWP), polarization beam-splitters (PBS), and a pair of detectors. The HWPs are responsible for selecting the measurement basis, as they rotate the initial polarization of the photons by an angle $\theta$. The input-output relations of a polarisation analyzer are
\begin{align}
&\hat{a}_{T_{A(B)}}=\hat{a}_{H_{A(B)}}\cos\theta_{A(B)}+\hat{a}_{V_{A(B)}}\sin\theta_{A(B)}\label{IOR1},\\
&\hat{a}_{R_{A(B)}}=-\hat{a}_{H_{A(B)}}\sin\theta_{A(B)}+\hat{a}_{V_{A(B)}}\cos\theta_{A(B)}\label{IOR2}.
\end{align}

The polarisation angles corresponding to measurements $\hat{\sigma_z}$, $\hat{\sigma_x}$, $(\hat{\sigma_x}+\hat{\sigma_z})/\sqrt{2}$, $(\hat{\sigma_x}-\hat{\sigma_z})/\sqrt{2}$ are $\theta_B^{(1)}=0$, $\theta_B^{(2)}=\pi/4$, $\theta_A^{(1)}=\pi/8$, $\theta_A^{(2)}=3\pi/8$. 

QBER for $\ket{\Phi^{+}}$ in terms of detection probabilities is defined as $Q=P_\mathrm{diff}\!\left(0,0\right)$. Similarly, the Bell parameter takes the form
\begin{equation}
\label{bell-p}
S =\left|E\left(0,
\frac{\pi}{8}\right)-E\left(0,\frac{3\pi}{8}\right)+E\left(\frac{\pi}{4},\frac{\pi}{8}\right)+E\left(\frac{\pi}{4},\frac{3\pi}{8}\right)\right|.
\end{equation}

At this stage, we have accounted for all imperfections arising from the polarization analyzer setups with SPAD detectors. Next, we consider the imperfections associated with different types of sources.

\subsection{Sources of light}
\label{section-4}

As already mentioned, we consider two types of sources: SPDC sources, which are the most commonly used in quantum-optical experiments, and novel QD sources of entangled photons. 

We assume that an SPDC source ideally produces a two-mode squeezed vacuum (TMSV) state \cite{Gerry_Knight_2004-1, Gerry_Knight_2004-2}, which in the Fock basis can be prepresented as
\begin{equation}
\label{spdc}
    \ket{\textit{SPDC}}= (\cosh{\xi})^{-1} \sum_{n=0}^{+\infty}\sqrt{n+1} (\tanh{\xi})^n \ket{\Phi_n},
\end{equation}
where $\xi$ is a squeezing parameter and 
\begin{equation}
    \ket{\Phi_n}= \frac{1}{\sqrt{n+1}}\sum_{m=0}^{n}(-1)^{m}
    \ket{n-m}_{H_{A}}\ket{m}_{V_{A}}\ket{m}_{H_{B}}\ket{n-m}_{V_{B}}.
\end{equation}

Alternatively to SPDC, we consider QD sources, explicitly including the practical imperfection due to FSS, which arises from the asymmetric fabrication of QDs \cite{rota2021}. To accurately represent the states produced by a QD source, we adopt the phase-modified Bell state model \cite{Huber2017-in, PhysRevLett.123.160501}:
\begin{equation}
\label{fss}
\ket{\tilde\Phi^{+}}=\frac{1}{\sqrt{2}}\,\left(\ket{0}_{H_A}\ket{0}_{H_B}+e^{-\frac{i \phi t}{\hbar}}\ket{1}_{V_A}\ket{1}_{V_B}\right),
\end{equation}
where $\phi$ represents the phase induced by the FSS during the generation of entangled photon states.

\subsection{Binning}
\label{section-5}

Note that for the entanglement-based BB84 protocol, the QBER is defined using only conclusive (coincident-click) outcomes, since the devices are trusted. Therefore, this subsection applies exclusively to the DI-QKD protocol analysis.

To account for imperfections in both detection setups and photon sources, we explicitly consider double-click events, including scenarios where both detectors on one side register simultaneous clicks ($P_{T_A T_B R_A}$, $P_{T_A T_B R_B}$, $P_{T_A R_B T_B}$, $P_{T_A R_B R_A}$) or when all detectors click simultaneously ($P_4$). Such events may arise from multiphoton emissions in SPDC sources or detector dark counts. Previous analyses often relied on the fair sampling assumption \cite{Semenov2009, Gumberidze2016}, treating registered detector clicks as representative of the complete measurement set. However, this assumption introduces a detection loophole \cite{Brunner2014} because events without detections are ignored. Non-conclusive events include no-click events ($P_0$), single-click events ($P_{T_A}, P_{T_B}, P_{R_A}, P_{R_B}$), and double-click events on a single side ($P_{T_A R_A}, P_{T_B R_B}$). Coincidence-click events ($P_{T_A T_B}, P_{T_A R_B}, P_{R_A T_B}, P_{R_A R_B}$) complete the full set of 16 possible detection outcomes, which must be appropriately binned into four logical outcomes: $\tilde{P}_{T_A T_B}, \tilde{P}_{T_A R_B}, \tilde{P}_{R_A T_B}, \tilde{P}_{R_A R_B}$.

One possible approach to binning detection events, suggested in Ref.~\cite{Vivoli2015}, is to assign the outcome $-1$ exclusively to transmitted (horizontal polarization) signals on each side and the outcome $+1$ to all other detection events. It optimizes the observed Bell violation $S$, leading to improved numerical results. However, in this work, we also adopt a standard binning method, in which we evenly distribute non-conclusive detection events among the four logical outcomes to simplify the calculation of the Bell parameter:
\begin{equation}
P_{\text{same}} - P_{\text{diff}} = \tilde{P}_{T_A T_B} + \tilde{P}_{R_A R_B} - \tilde{P}_{T_A R_B} - \tilde{P}_{R_A T_B},
\end{equation}
with the binning strategy
\begin{gather}
\label{binning}
\tilde{P}_{T_A, T_B} = P_{T_A, T_B} + \frac{1}{4}\left(P_{T_A,R_A}+P_{T_B,R_B}+P_{0}+P_{4}\right)+ \frac{1}{2}\left(P_{T_A}+P_{T_B}+P_{T_A, T_B, R_A}+P_{T_A, T_B, R_B}\right), \\
\tilde{P}_{T_A, R_B} = P_{T_A, R_B} + \frac{1}{4}\left(P_{T_A,R_A}+P_{T_B,R_B}+P_{0}+P_{4}\right)+ \frac{1}{2}\left(P_{T_A}+P_{R_B}+P_{T_A, R_B, T_B}+P_{T_A, R_B, R_A}\right), \\
\tilde{P}_{R_A, T_B} = P_{R_A, T_B} + \frac{1}{4}\left(P_{T_A,R_A}+P_{T_B,R_B}+P_{0}+P_{4}\right)+ \frac{1}{2}\left(P_{R_A} + P_{T_B} + P_{R_A, T_B, T_A}+P_{R_A, T_B, R_B}\right), \\
\tilde{P}_{R_A, R_B} = P_{R_A, R_B} + \frac{1}{4}\left(P_{T_A,R_A}+P_{T_B,R_B}+P_{0}+P_{4}\right) + \frac{1}{2}\left(P_{R_A}+P_{R_B}+P_{R_A, R_B, T_A}+P_{R_A, R_B, T_B}\right).
\end{gather}

This binning preserves the analytic relation \(S = 2\sqrt{2}(1 - 2Q)\) for QD sources subject to depolarizing noise. Additionally, it yields significantly lower QBER for SPDC sources under realistic experimental conditions. The respective analytical expressions for the alternative binning \cite{Vivoli2015} are provided in the Appendix 1. 

\section{Performance of protocols}
\label{resultsDI}

\subsection{SPDC sources}

In the case of SPDC sources, it is convenient to move to phase space to calculate detection probabilities \eqref{ProbabilitySquash}. Therefore, we switch to the Glauber-Sudarshan representation of the density matrix, furthermore, it is better to work with characteristic function (Fourier transform of the Glauber-Sudarshan function) due to its positivity in the whole space. The choice of Glauber- Sudarshan function is not obligatory, other alternatives can be used: Qusimi or Wigner functions. For simplicity, we neglect depolarization in the channel when analyzing this source.

\begin{figure}
\centering 
\begin{tikzpicture}
\node (img1){\includegraphics[width=0.7\columnwidth]{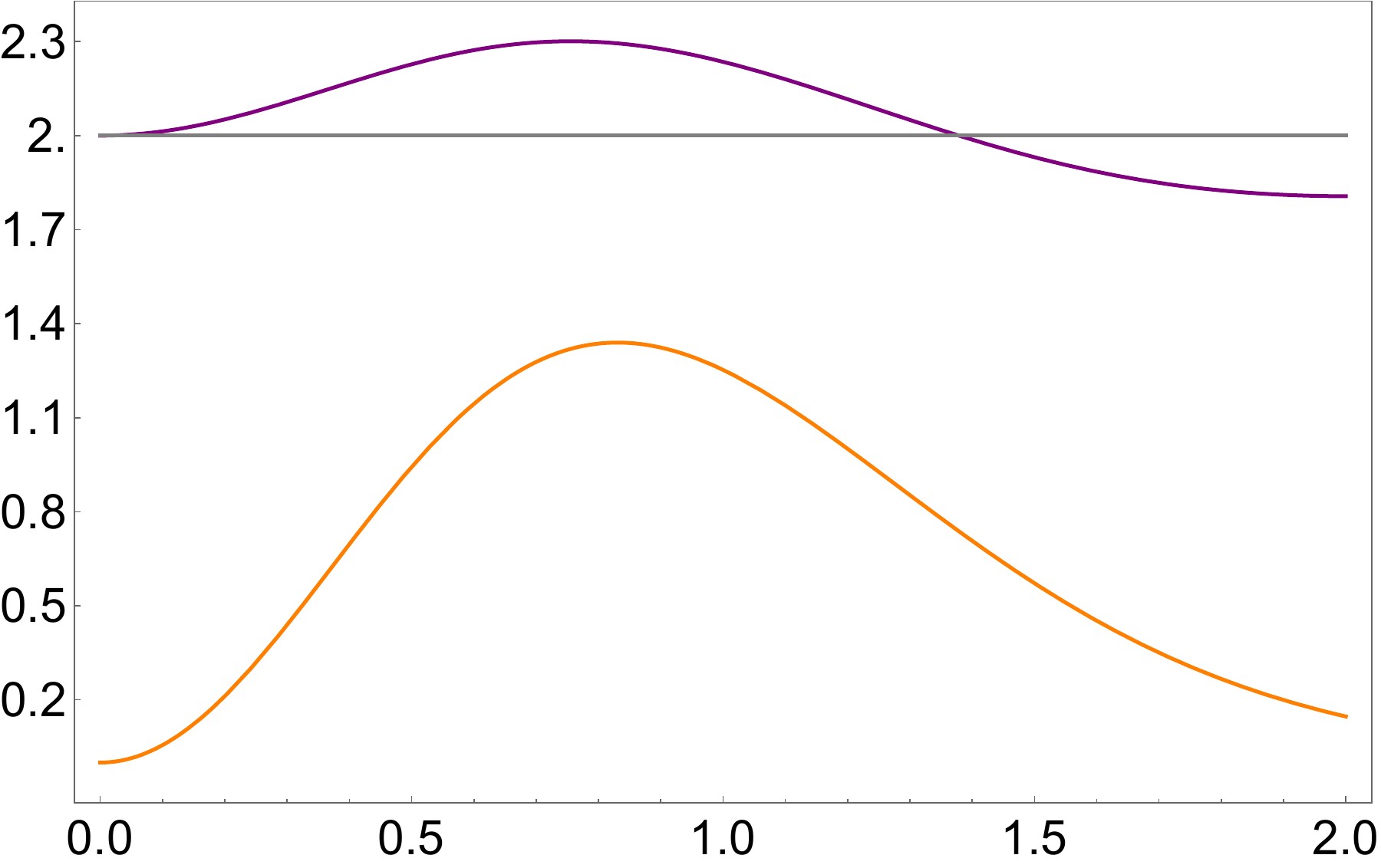}};
\node[left=of img1, node distance=0cm,rotate=90, yshift=-0.5cm,xshift=1.5cm] {$\mathcal{S}, \textit{Bell parameter}$};
\node[below=of img1, node distance=0cm, yshift=0.9cm,xshift=0.2cm] {$\xi, \textit{squeezing parameter}$};
\node[above=of img1, node distance=0cm, rotate=30, yshift=-5.2cm,xshift=-2.8cm] {$\textit{QBER}$};
\node[in=of img1, node distance=0cm, yshift=0.9cm,xshift=3.8cm]{$\eta =1$};
\node[in=of img1, node distance=0cm, yshift=.4cm,xshift=3.8cm]{$\nu = 0 $};
\node[in=of img1, node distance=0cm, yshift=-1.6cm,xshift=-0.3cm]
{\begin{minipage}{.26\textwidth}
 \begin{center}
{\includegraphics[width=\columnwidth]{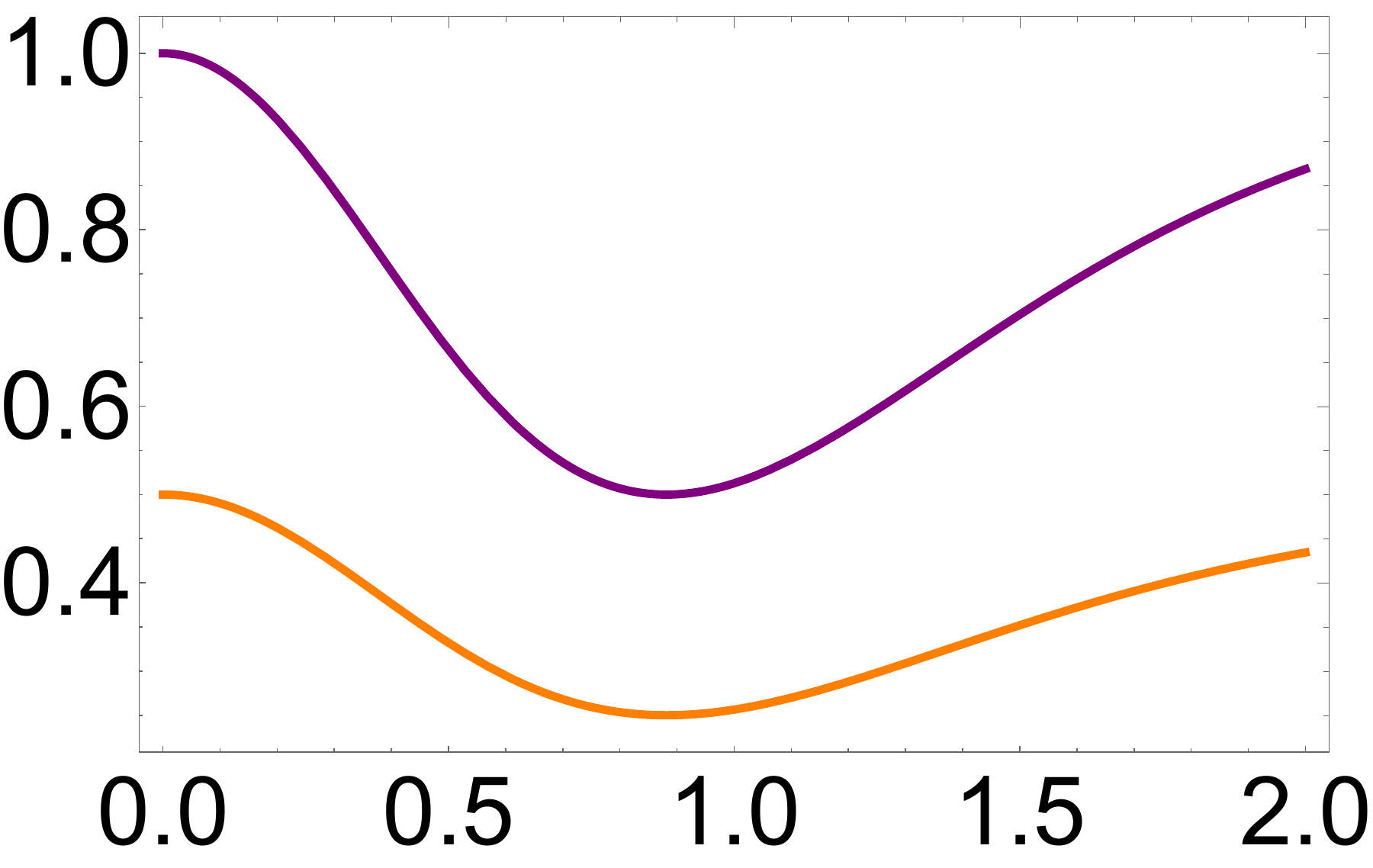}}
 \end{center}
 \end{minipage}};
\end{tikzpicture}
\caption{\label{S-Q-s-spdc} The Bell parameter \eqref{bell-p} is plotted as a function of the squeezing parameter $\xi$ for SPDC source with the standard binning in Sec.~\ref{section-5} (orange line) and with alternative binning \cite{Vivoli2015} (violet line). The inset illustrates the dependence of QBER on $\xi$ with the same colour code. In the case of ideal pair generation, the optimal measurements $\theta_{A}^{(1)} = \pi/8$, $\theta_{A}^{(2)} = 3\pi/8$, $\theta_{B}^{(1)} = 0$, and $\theta_{B}^{(2)} = \pi/4$ lead to the maximum CHSH Bell inequality violation \eqref{bell}. However, for SPDC sources, vacuum and multiphoton events suppress the violation under the current binning strategy. Results with an alternative binning strategy \cite{Vivoli2015} yield Bell inequality violations up to $S = 2.30083$ under optimized measurement angles ($\theta_{A}^{(1)} = 0.661$, $\theta_{A}^{(2)} = 1.248$, $\theta_{B}^{(1)} = 2.525$, $\theta_{B}^{(2)} = 3.112$) and a squeezing parameter of $\xi = 0.755$. Both main plot and inset correspond to DI-QKD protocol analysis.}

\end{figure}
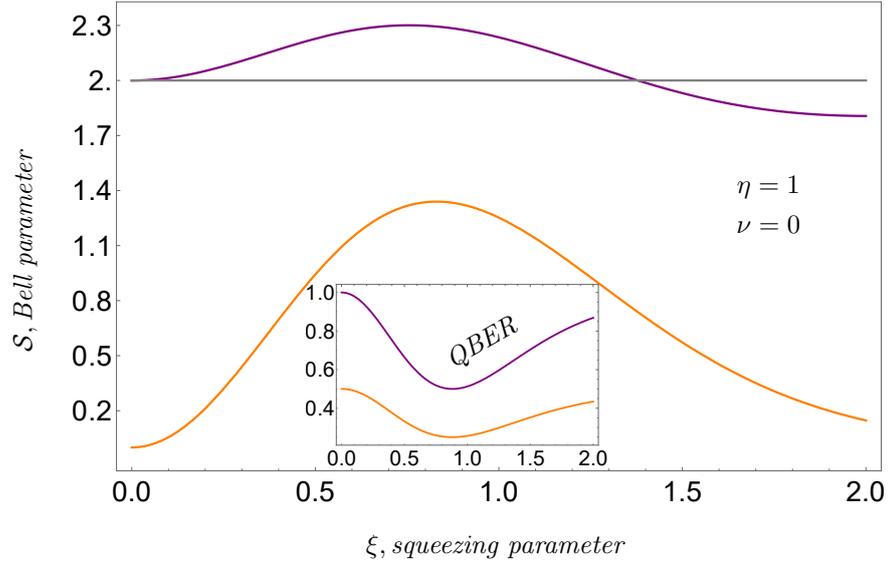

The characteristic function $\Phi$ of the detected light \cite{Semenov2010, Gumberidze2016, Tsujimoto2018} is given by
\begin{gather}
\displaystyle{\Phi\left(\beta_{T_A}, \beta_{R_A}, \beta_{T_B}, \beta_{R_B}\right)=}
\label{charact_func}\\
\nonumber
\displaystyle{=\exp \left[-\frac{\tanh^2\xi}{1-\tanh^2\xi}\Big(\left|\beta_{T_A}\right|^2+\left|\beta_{R_A}\right|^2+\left|\beta_{T_B}\right|^2+\left|\beta_{R_B}\right|^2\Big)\right]\times} \\
\nonumber
\displaystyle{\exp\left[-\frac{\tanh\xi}{1-\tanh^2\xi}\left(\beta_{T_A}\beta_{T_B}+\beta_{T_A}^{*}\beta_{T_B}^{*}+\beta_{R_A}\beta_{R_B}+\beta_{R_A}^{*}\beta_{R_B}^{*}\right)\sin{\left(\theta_A-\theta_B\right)}\right]\times}
\\
\displaystyle{\exp\left[-\frac{\tanh\xi}{1-\tanh^2\xi}\left(\beta_{T_A}\beta_{R_B}+\beta_{T_A}^{*}\beta_{R_B}^{*}-\beta_{R_A}\beta_{T_B}-\beta_{R_A}^{*}\beta_{T_B}^{*}\right)\cos{\left(\theta_A-\theta_B\right)}\right]},
\nonumber
\end{gather}
where the output modes of the polarization analyzers, 
$\vec{\beta} = \{\beta_{T_A}, \beta_{R_A}, \beta_{T_B}, \beta_{R_B}\}$,  
are related to the corresponding input modes,  
$\vec{\beta} = \{\beta_{H_A}, \beta_{V_A}, \beta_{H_B}, \beta_{V_B}\}$,  
via the input–output relations:
\begin{align}
\beta_{H_{A(B)}} &= \beta_{T_{A(B)}} \cos \theta_{A(B)} - \beta_{R_{A(B)}} \sin \theta_{A(B)}, \label{eq:C5a}\\
\beta_{V_{A(B)}} &= \beta_{T_{A(B)}} \sin \theta_{A(B)} + \beta_{R_{A(B)}} \cos \theta_{A(B)}. \label{eq:C5b}
\end{align}
More details of the calculations can be found in Appendix 2.

Here we present the final results for joint probabilities: 
\begin{gather}
\label{p-j-res}
    \tilde{P}_{T_A T_B} = \tilde{P}_{R_A R_B} = \frac{1}{4} + \frac{e^{-2\nu}}{2 \eta^2}\left\{ \frac{1}{\zeta^2 - \gamma^2} - \frac{1}{\zeta^2 - \lambda^2} \right\}, \\
    \tilde{P}_{T_A R_B} = \tilde{P}_{R_A T_B} = \frac{1}{4} + \frac{e^{-2\nu}}{2 \eta^2}\left\{ \frac{1}{\zeta^2 - \lambda^2} - \frac{1}{\zeta^2 - \gamma^2} \right\},
\end{gather}
and
\[P_\text{same} - P_\text{diff} = \frac{2e^{-2\nu}}{\eta^2}\left\{ \frac{1}{\zeta^2 - \gamma^2} - \frac{1}{\zeta^2 - \lambda^2} \right\}, \]
where the following notations were used

\begin{gather}
\label{alpha}
\zeta = \frac{\tanh^2{\xi}}{1-\tanh^2{\xi}} + \frac{1}{\eta},\\
\label{beta}
\gamma = \frac{\tanh{\xi}}{1-\tanh^2{\xi}} \sin{\left(\theta_\mathrm{A}-\theta_\mathrm{B}\right)},\\
\label{gamma}
\lambda = \frac{\tanh{\xi}}{1-\tanh^2{\xi}} \cos{\left(\theta_\mathrm{A}-\theta_\mathrm{B}\right)}. 
\end{gather}
We calculate the Bell parameter by substituting the joint probabilities \eqref{p-j-res} in equation \eqref{correlation} to determine the correlations, which are then plugged into equation \eqref{bell}.

As shown in Fig.~\ref{S-Q-s-spdc}, the Bell parameter remains below the classical bound for the symmetric binning strategy, described in Subsection~\ref{section-5}, even though the measurement angles that maximize CHSH violation for an ideal Bell pair remain optimal for SPDC sources with vacuum and multiphoton components. Consequently, the Devetak-Winter key rate bound is negative, indicating that secure QKD is not possible under these conditions. On the other hand, an alternative binning strategy \cite{Vivoli2015, Tsujimoto2018} with optimized angles can reach the Bell parameter of \(\ S\approx 2.3 \); however, it comes at the cost of a significantly increased QBER as shown in the inset of Fig.~\ref{S-Q-s-spdc}. Nevertheless, the violation remains insufficient for secure DI-QKD in agreement with previous findings~\cite{Vivoli2015}. Additionally, consistent with ~\cite{Scarani2008, Kravtsov_2023} and our analysis, vacuum effects at low pump power and multi-pair errors at higher power make a bare SPDC source unsuitable for secure entanglement-based BB84. However, including the squashing model improves the results by showing parameter ranges where secure keys can be generated, emphasizing the importance of realistic detector modeling. Although our analysis focuses on the estimation of the Bell parameter and QBER under realistic noise, a full finite-size key-rate evaluation lies beyond the scope of this work, aimed at establishing the bounds on the protocol feasibility already in the asymptotic regime. Future research may also address protocol modifications, such as incorporating decoy-state methods or refining security models to better capture imperfections introduced by practical photon sources.

\subsection{QD sources}

QD sources typically exhibit a drastically lower probability of multiphoton emissions compared to SPDC sources. However, they are susceptible to FSS, which arises from asymmetries in the QD structure during the growth process. These asymmetries can lead to variations in the confinement potential, resulting in FSS \cite{Basset2021, Schimpf2021}. This splitting introduces an additional phase in the entangled photon pairs, necessitating an evaluation of its impact on the secure key rate. Furthermore, we consider the presence of white noise in this scenario, leading to the final state described as
\begin{equation}
\label{qd-state}
    \hat{\rho}(p)= p \ket{\tilde\Phi^{+}}\bra{\tilde\Phi^{+}}+\frac{1-p}{4}\;\hat{\mathrm{I}}\otimes\hat{\mathrm{I}}.
\end{equation}

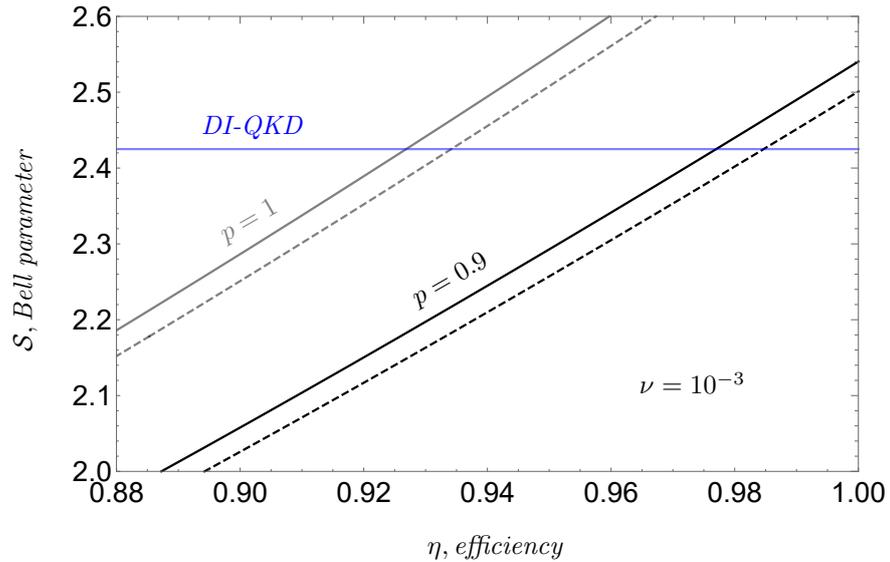
\begin{figure}
\centering 
\begin{tikzpicture}
\node (img1){\includegraphics[width=0.7\columnwidth]{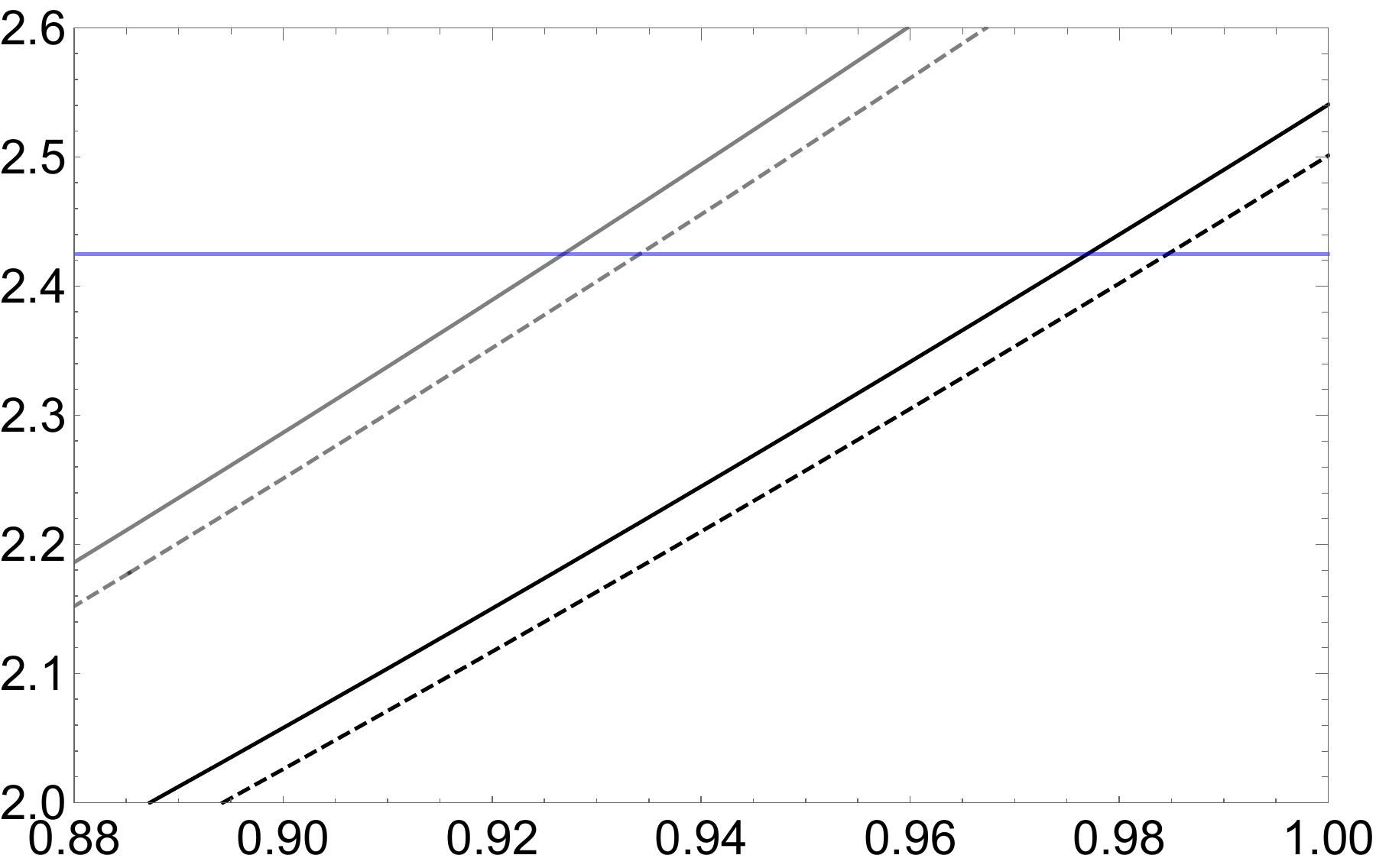}};
\node[left=of img1, node distance=0cm,rotate=90, yshift=-0.5cm,xshift=1.5cm] {$\mathcal{S}, \textit{Bell parameter}$};
\node[below=of img1, node distance=0cm, yshift=0.9cm,xshift=0.2cm] {$\mathcal{\eta}, \textit{efficiency}$};
\node[in=of img1, node distance=0cm, rotate=30, yshift=-0.1cm,xshift=-0.5cm]{$p = 0.9$};
\node[in=of img1, node distance=0cm, rotate=30, yshift=1.9cm,xshift=-2.4cm]{\color{gray}$p = 1$};
\node[in=of img1, node distance=0cm, yshift=-1.7cm,xshift=2.8cm]{$\nu = 10^{-3} $};
\node[above=of img1, node distance=0cm, rotate=0,yshift=-3.1cm,xshift=-3.cm] {{\color{blue}$\textit{DI-QKD}$}};
\end{tikzpicture}
\caption{\label{S-eta-qd} The Bell parameter \eqref{bell-p} is plotted as a function of the detection efficiency $\eta$ for a QD source without (solid line) and with (dashed line) the effect of FSS. The results demonstrate that the Bell-parameter violation remains sufficient for secure DI-QKD, even in the presence of FSS and imperfect detection, under a dark count rate of $\nu=10^{-3}$ and an initial state survival probability of $p=0.9$ after depolarisation in the channel. The horizontal blue line indicates the critical minimum Bell parameter required for the DI-QKD secure protocol with the given parameters; values above this threshold (upper half) correspond to the secure DI-QKD regime. For comparison, we also plot the scenario in which polarisation effects in the channel are neglected ($p = 1$, gray lines), which reduces the minimum detection efficiency required for secure DI-QKD.}
\end{figure}

\begin{figure}
\centering 
\begin{tikzpicture}
\node (img1){\includegraphics[width=0.7\columnwidth]{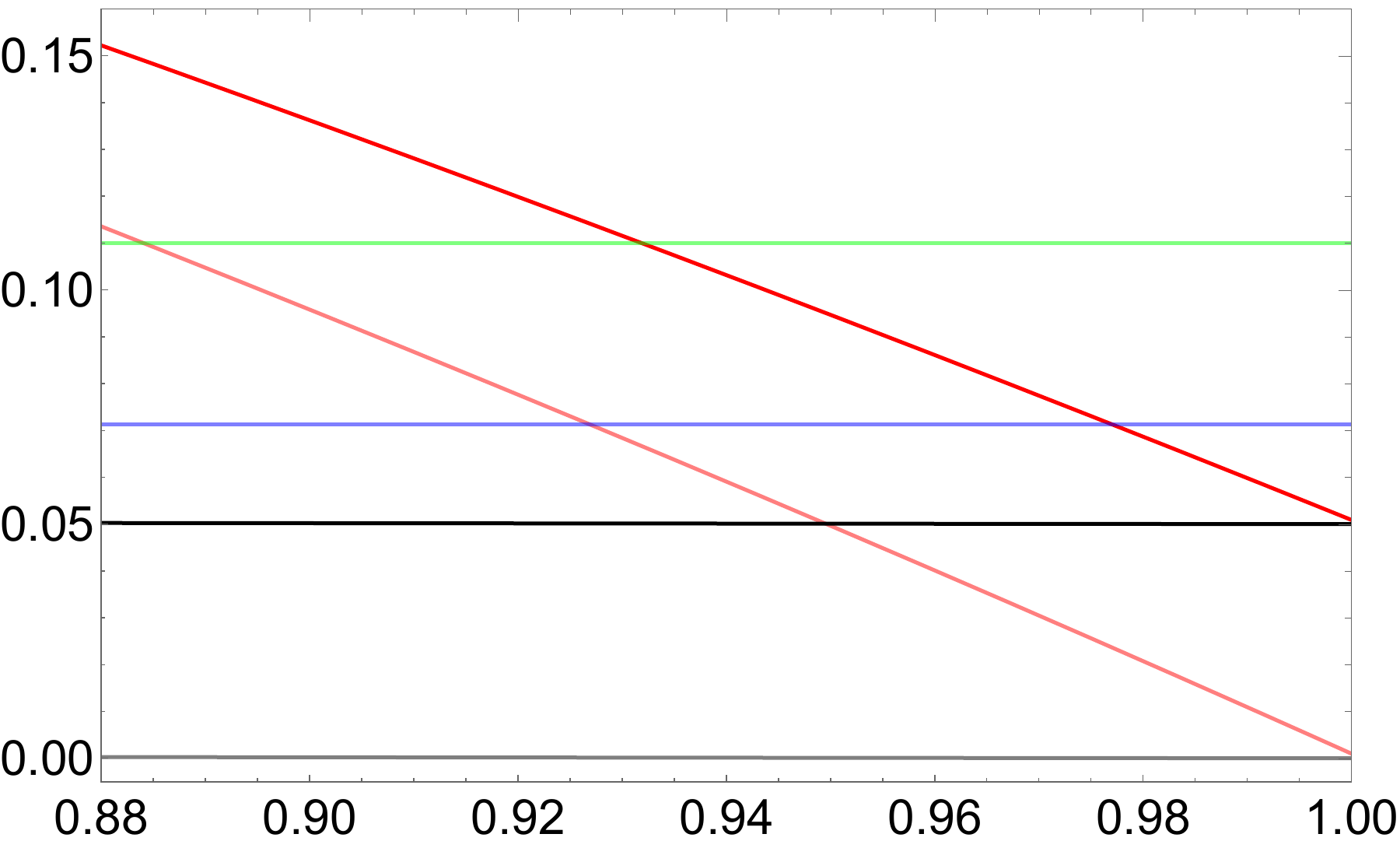}};
\node[left=of img1, node distance=0cm,rotate=90, yshift=-0.6cm,xshift=0.9cm] {$\mathcal{Q}, \textit{QBER}$};
\node[below=of img1, node distance=0cm, yshift=0.9cm,xshift=0.2cm] {$\mathcal{\eta}, \textit{efficiency}$};
\node[in=of img1, node distance=0cm, yshift=-.5cm,xshift=-2.5cm]{$p = 0.9$};
\node[in=of img1, node distance=0cm, yshift=-2.3cm,xshift=-1.cm]{\color{gray}$p = 1$};
\node[in=of img1, node distance=0cm, rotate = -22, yshift=1.4cm,xshift=1.8cm]{\color{red}$p = 0.9$};
\node[in=of img1, node distance=0cm, rotate = -25, yshift=-.cm,xshift=3.7cm]{\color{red!60!white}$p = 1$};
\node[in=of img1, node distance=0cm, yshift=2.3cm,xshift=3.5cm]{$\nu = 10^{-3} $};
\node[above=of img1, node distance=0cm, rotate=0,yshift=-3.cm,xshift=1.5cm] {{\color{green}$\textit{BB84}$}};
\node[above=of img1, node distance=0cm, rotate=0,yshift=-4.4cm,xshift=-3.3cm] {{\color{blue}$\textit{DI-QKD}$}};
\end{tikzpicture}
\caption{\label{Q-eta-qd} The DI-QKD QBER (red line) and the entanglement-based BB$84$ QBER (black line) versus the detection efficiency $\eta$ for the QD source with and without the effect of FSS (both overlap, indicating that the impact of FSS completely cancels out in QBER for both types of protocols). The analysis assumes imperfect detection with a dark-count probability of $\nu=10^{-3}$ and an initial photon-pair survival probability $p=0.9$ after depolarization in the quantum channel. Horizontal green and blue lines mark the maximal tolerable QBER for secure BB84 and secure DI-QKD, respectively; values below these lines correspond to the secure operating regime. For comparison, we also plot the case without channel depolarisation, which lowers the detection-efficiency requirement for DI-QKD. Note that for BB84 the QBER values remain well below the corresponding threshold across the plotted range; in the absence of depolarisation the protocol becomes fully secure, as the QD state \eqref{qd-state} approaches an ideal Bell state. }
\end{figure}
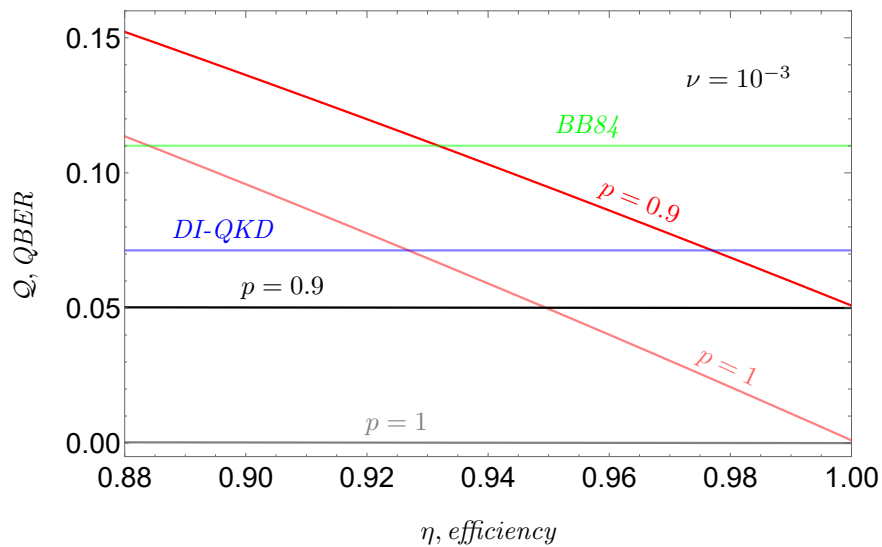

\begin{figure}
\centering 
\begin{tikzpicture}
\node (img1){\includegraphics[width=0.7\columnwidth]{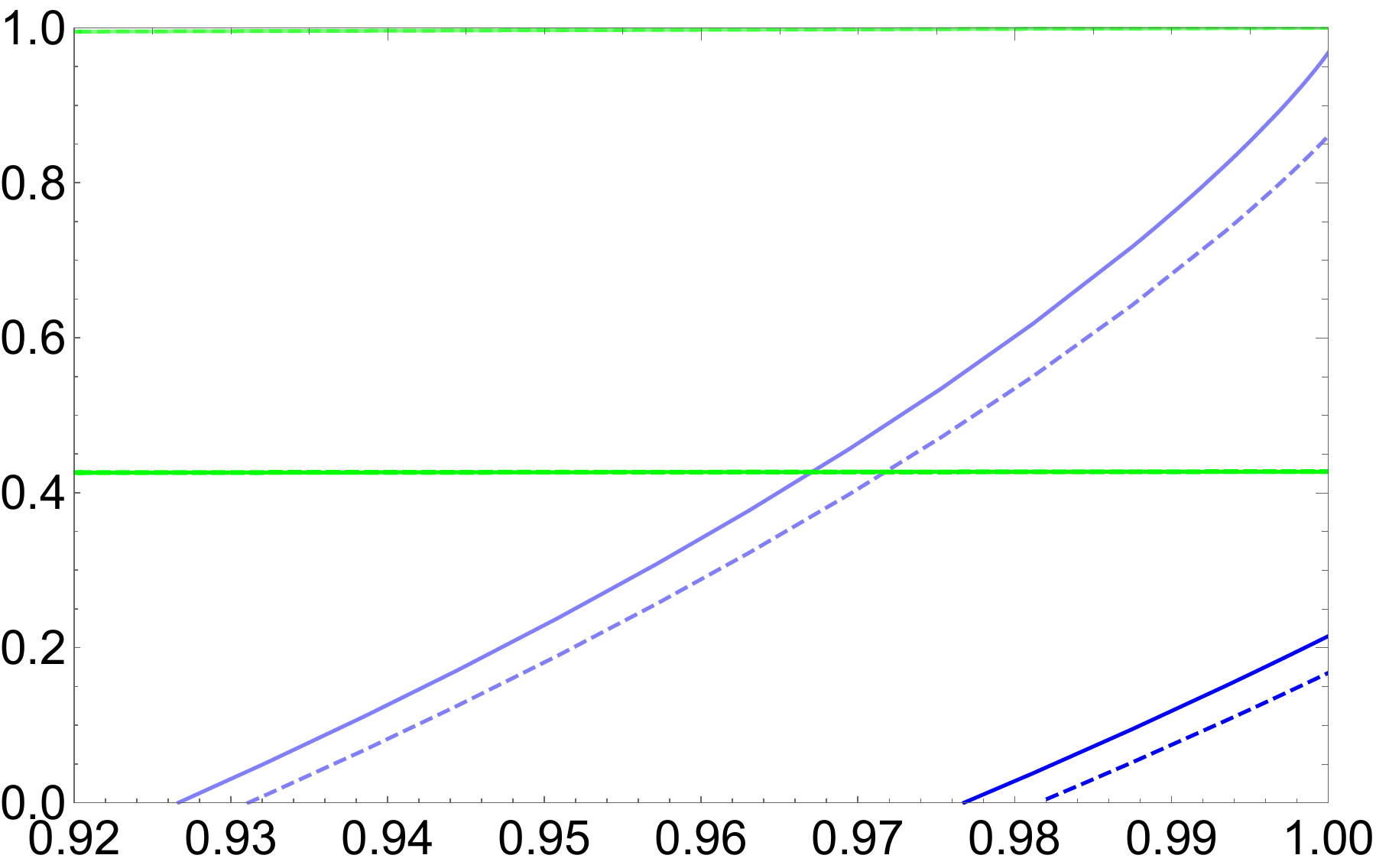}};
\node[left=of img1, node distance=0cm,rotate=90, yshift=-0.7cm,xshift=1.8cm] {$r_{\textit{DW}}, \textit{D.-W. key rate}$};
\node[below=of img1, node distance=0cm, yshift=0.9cm,xshift=0.2cm] {$\mathcal{\eta}, \textit{efficiency}$};
\node[in=of img1, node distance=0cm, rotate=25, 
yshift=-3.2cm,xshift=2.4cm]{{\color{blue}$\textit{DI-QKD} \,(p = 0.9)$}};
\node[in=of img1, node distance=0cm, rotate = 25, yshift=-4.1cm,xshift=2.8cm]{\color{blue}$\textit{FSS}$};
\node[in=of img1, node distance=0cm, yshift=1.6cm,xshift=-2.4cm]{$\nu = 10^{-3} $};
\node[above=of img1, node distance=0cm, rotate=27,yshift=-4.8cm,xshift=-4.3cm] {{\color{blue!60!white}$\textit{DI-QKD} \,(p = 1)$}};
\node[above=of img1, node distance=0cm, rotate=30,yshift=-5.6cm,xshift=-3.2cm] {{\color{blue!60!white}$\textit{DI-QKD}(\textit{FSS}, p=1)$}};
\node[above=of img1, node distance=0cm, rotate=0,yshift=-2.1cm,xshift=.cm] {{\color{green!60!white}$\textit{BB84}\,(p = 1)$}};
\node[above=of img1, node distance=0cm, rotate=0,yshift=-4.8cm,xshift=-2.3cm] {{\color{green}$\textit{BB84}\,(p = 0.9)$}};
\end{tikzpicture}
\caption{\label{r-eta} The extractable Devetak–Winter key rates against collective attacks for DI-QKD (blue) and BB84 (green) are plotted versus the detection efficiency $\eta$ for a QD source. For DI-QKD, the curve including FSS is shown dashed; for BB84, FSS has no impact on security and a single solid curve suffices. The analysis includes imperfect detection with a dark-count probability $\nu=10^{-3}$ and an initial-state survival probability $p=0.9$ after depolarisation in the channel. The plot highlights that DI-QKD requires highly efficient detectors, posing an experimental challenge. For comparison, semi-transparent lines depict the scenario without channel depolarisation, which shifts the DI-QKD security threshold to lower detection efficiencies. Note that BB84’s security exhibits only a weak dependence on detection efficiency and is shown primarily for comparison.}
\end{figure}
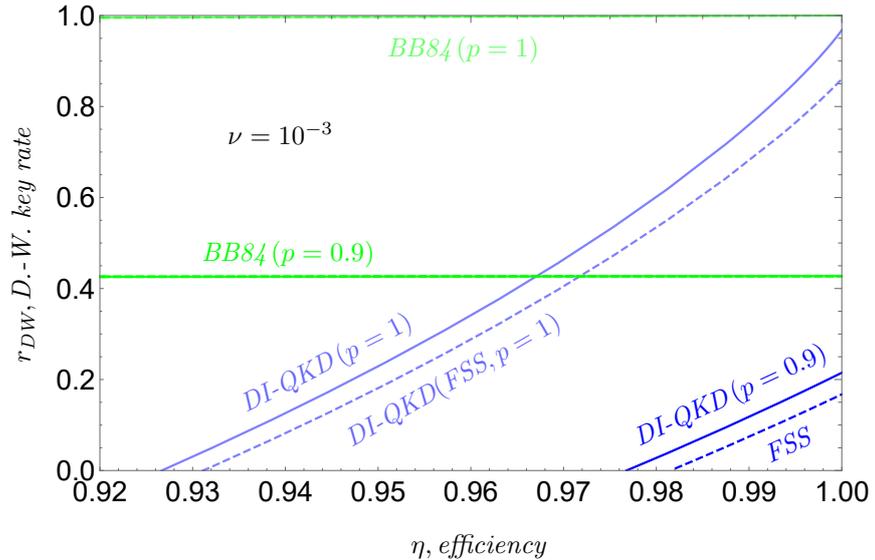

For DI-QKD resulting expressions for the Bell parameter and QBER in the presence of FSS are given by:

\begin{gather}
S =\frac{e^{-\frac{i \Phi t}{\hbar}-2n}\left(1 + e^{\frac{i \Phi t}{\hbar}}\right)^2 p \,\eta^2}{\sqrt{2}}, \\
Q= \frac{1}{2}\left(1-e^{-2 \nu} p\, \eta^2\right).
\end{gather}
Note that FSS affects only the Bell parameter; the QBER remains unchanged because the FSS-dependent terms cancel out in its expression. For comparison, in the absence of FSS, the Bell parameter simplifies to \(2\sqrt{2}\, e^{-2\,\nu} \, p \, \eta^2\) , while the QBER retains the same form. As discussed earlier, the correlation structure \(S = 2\sqrt{2}(1 - 2Q)\) holds only in the absence of FSS; once FSS is introduced, this relation no longer applies. 

For entanglement-based BB$84$, the QBER is defined using only conclusive detection events. Similarly, the effect of FSS cancels out, resulting in identical QBER and thus identical key rates for the cases with and without FSS.

Figures \ref{S-eta-qd} and \ref{Q-eta-qd} depict the Bell parameter violation and QBER for high detection efficiencies $\eta$, with horizontal lines marking critical thresholds for QKD protocols. Despite the presence of FSS, QD sources remain viable for both DI-QKD and entanglement-based BB84. Figure~\ref{r-eta} presents secure key rates as a function of detection efficiency, using parameters $p=0.9$ \cite{Grasselli2023boostingdevice}, $\nu=10^{-3}$ typical for SPAD \cite{Slussarenko}, and FSS $-\frac{i \phi t}{\hbar}=0.25$, typical for GaAs QD sources \cite{Huber2017-in, article}. The results show that DI-QKD requires high Bell parameter values and detection efficiencies approaching $\eta \approx 1$. 


Figures~\ref{S-eta-qd} and \ref{Q-eta-qd} show the Bell parameter violation and QBER as functions of the detection efficiency $\eta$, with horizontal lines marking the critical thresholds for QKD protocols. Despite the presence of FSS, QD sources remain viable for both DI-QKD and entanglement-based BB84. Figure~\ref{r-eta} further presents the corresponding secret key rates, calculated using parameters $p=0.9$ \cite{Grasselli2023boostingdevice}, $\nu=10^{-3}$ - dark count rate per detection window typical for state-of-the-art SPAD detectors \cite{Slussarenko}, and FSS $-\frac{i \phi t}{\hbar}=0.25$, typical for GaAs QD sources \cite{Huber2017-in, article}.

The results indicate that DI-QKD requires very high detection efficiencies, approaching the ideal limit $\eta \to 1$, in order to achieve a sufficient Bell violation. By contrast, the entanglement-based BB84 protocol is essentially independent of efficiency, yielding a nearly flat key-rate curve. For this reason, the BB84 results are included mainly for reference and comparison, while the feasibility of DI-QKD under realistic detector parameters remains the central focus of our analysis. As in all figures, we also overlay the no-depolarisation scenario using semi-transparent lines to isolate the pure impact of FSS and to show the shift toward lower detection-efficiency requirements for secure DI-QKD (i.e., an expanded feasible range).

\section{Conclusions and Outlook}
\label{concls}

We investigated the practical performance of entanglement-based QKD protocols — device-independent and BB$84$ — by comparing prominent entangled photon pair sources, specifically SPDC and QD, within a unified analytical framework. Rather than tailoring security proofs to each source individually, our analysis focuses on estimating the Bell parameter and QBER under realistic detection conditions in order to compare the performance of the different sources in entanglement-based QKD.

We explicitly include practical imperfections such as limited detector (yet high) efficiency and dark counts, alongside critical source-specific issues—vacuum and multiphoton emissions for SPDC sources, and FSS for QD sources. Despite the detrimental impact of FSS, QD sources still achieve sufficient Bell inequality violations and positive Devetak–Winter key rates under realistic experimental conditions. In contrast, SPDC sources fail to surpass the classical Bell bound under standard measurement angles and symmetric binning, even assuming ideal detection, due to vacuum-dominated emissions at low gain and increased QBER from multipair emissions at higher gain. Although optimized measurement angles and asymmetric binning strategies can improve the Bell parameter for SPDC, these adjustments alone do not yield positive key rates. Future secure implementation of SPDC-based QKD thus requires additional enhancements—such as heralding or decoy-state approaches—whose security must be carefully examined with adapted proofs. Exploring these methods, along with finite-size effects, remains a promising direction for further research.

Furthermore, while we have computed Devetak–Winter key rates following the approach in Refs.~\cite{Acin2007PRL, Pironio_2009}, there is potential for improved rates through alternative DI-QKD modifications, such as those leveraging Jordan’s lemma \cite{Primaatmaja2023securityofdevice} or employing semi-definite programming methods \cite{Araujo2023quantumkey, Brown_2024, Brown2021-ug}.

\ack{The authors acknowledge valuable advice from prof. A. Acin and his group at ICFO, Barcelona. M.G. and V.U. acknowledge the project $8$C$22003$ (QD-E-QKD) of the Czech MEYS and project $21$-$44815$L of the Czech Science Foundation, V.U. acknowledges the project CZ.$02.01.01/00/22$\_$008/0004649$ (QUEENTEC) of the Czech MEYS.}

\funding{Czech MEYS (CZ.$02.01.01/00/22$\_$008/0004649$); Czech MEYS ($8$C$22003$); Czech Science Foundation ($21$-$44815$L).}

\section*{Disclosures} 

The authors declare no conflicts of interest.

\data{Data underlying the results presented in this paper are not publicly available at this time but may be obtained from the authors upon reasonable request.}

\suppdata{See Supplemental material for supporting content.}

\bibliography{references}
\bibliographystyle{unsrt}

\end{document}